\documentclass[aps,prb,superscriptaddress,showpacs]{revtex4-1}
\usepackage{graphicx,amsmath,amssymb,xspace,epsfig,float,multirow,subfigure,tabularx}

\pdfoutput=1

\begin{document}

\title{Dynamical instability of the electric transport in strongly
fluctuating superconductors.}
\author{Lei Qiao}
\affiliation{School of Physics, Peking University, Beijing 100871, \textit{China}}
\affiliation{Collaborative Innovation Center of Quantum Matter, Beijing, China}
\author{Dingping Li}
\email{lidp@pku.edu.cn}
\affiliation{School of Physics, Peking University, Beijing 100871, \textit{China}}
\affiliation{Collaborative Innovation Center of Quantum Matter, Beijing, China}
\email{lidp@pku.edu.cn}
\author{Baruch Rosenstein}
\email{baruchro@hotmail.com}
\affiliation{Electrophysics Department, National Chiao Tung University, Hsinchu 30050,
\textit{Taiwan, R. O. C}}
\affiliation{Physics Department, Bar-Ilan University, 52900 Ramat-Gan, Israel}
\date{\today }

\begin{abstract}
Theory of the influence of the thermal fluctuations on the electric
transport beyond linear response in superconductors is developed within the
framework of the time dependent Ginzburg - Landau approach. The I - V curve
is calculated using the dynamical self - consistent gaussian approximation.
Under certain conditions it exhibits a reentrant behaviour acquiring an S -
shape form. The unstable region below a critical temperature $T^{\ast }$ is
determined for arbitrary dimensionality ($D=1,2,3$) of the thermal
fluctuations. The results are applied to analyse the transport data on
nanowires and several classes of 2D superconductors: metallic thin films,
layered and atomically thick novel  materials.
\end{abstract}

\maketitle

\section{Introduction}

In most electrical transport phenomena in condensed matter the current in a
conductor is a monotonic function of the applied voltage. Moreover at small
current densities the I-V curve is nearly linear (see the dark green line in
Fig.1 representing a normal metal), so that only the linear response theory%
\cite{Mahan} is generally needed to describe the transport via conductivity $%
\sigma _{n}$. However in certain types of materials the linearity does not
extend to higher current densities. In superconductors close to the normal
state (when temperature for example is just below critical, see the purple
curve in Fig.1), at small currents the I-V curve slope $\sigma $ is very
large $\sigma >>\sigma _{n}$, however at higher currents it diminishes and
then smoothly approaches the normal line.

It turns out that under certain conditions (for example at yet lower
temperatures, the solid cyan curve in Fig.1), the initial slope is even
steeper and moreover at certain current density the differential resistivity
becomes negative signalling a dynamical instability.

This possibility was envisioned theoretically by Gorkov\cite{Gorkov} and
Masker, Marcelja and Parks\cite{Parks}, before strongly fluctuating
superconductors like the high $T_{c}$ cuprates were discovered. The
arguments required strong fluctuations that enhance conductivity of a metal,
beyond the parameter range in which the coherent condensate is not formed.

The theory in the one - dimensional geometry was discussed in a
comprehensive paper by Tucker and Halperin\cite{Halperin71}. Different
versions of the dynamical Hartree - Fock approximation were critically
compared. The focus on wires (one dimensions, 1D) was justified, since low $%
T_{c}$ superconductors have very small Ginzburg number $Gi$ and the
fluctuations are detectable only when the dimensionality is reduced (or
strong magnetic fields applied). The Tucker and Halperin conclusion was that
the approximation is probably inapplicable for currents for which
differential resistivity is negative, but qualitatively the phenomenon
should be observable in 1D. Later several experiments indeed appeared both
in 1D (thin metallic nanowires)\cite{1Dexperiment} and in 2D both in thin
metallic films\cite{Ivanchenko} and layered high $T_{c}$ materials\cite%
{Xiao99,Yeshurun}, that have a much larger Ginzburg number, so that thermal
fluctuations in them are much easier to observe. Moreover recently purely 2D
superconductors (with thickness of just one or very few unit cells) appeared%
\cite{Lin15,Wangscience,WangNbSe} and similar phenomenon was observed.

It is important to note that, due to experimental reasons, only in the first
two experiments\cite{1Dexperiment,Ivanchenko} the voltage drive was used, so
that the full I-V curve including the ``unstable" parts was observed. In rest
of experiments the current drive was employed, so one observed that at
certain current the voltage ``jumped" over the unstable state. Many more
experiments observing instability (with jumps due to the current drive) were
performed in superconducting films\cite{Xiao99,Huebener,Tsuei} and wires
under strong magnetic field. In the presence of magnetic field in type II
superconductors the dynamical problem becomes more complex due to effects of
the vortex pinning and theoretical explanations invoke thermal transport
(hot spots\cite{Mints}). The experiment on 1D nanowires was qualitatively
explained\cite{Vodolazov}, using dynamics of the condensate, rather then
utilizing Tucker-Halperin theory. As was noted early on\cite%
{Gorkov,Ivanchenko,Mints}, the dynamical instability, is firmly established,
can leads to dynamical phase separation patterns and other phenomena and
applications.

In this paper we revisit and expand the self - consistent theory of the
nonlinear response in superconductors and demonstrate that the old and the
new experiments on the dynamical instability can in fact be explained by it,
not just qualitatively, but quantitatively. The conditions for the
instability are derived in D=1,2 and even D=3 (in which case these are
almost impossible to observe even for the most ``fluctuating" materials). It
seems that a covariant version of the dynamical gaussian approximation\cite%
{Kovner} in the framework of the Ginzburg - Landau phenomenological approach%
\cite{Rosenstein} is precise and universal enough to quantitatively describe
the phenomenon \textit{including} the unstable regions.

A qualitative argument ensuring the emergence of the dynamical instability
for superconductors that posses large enough thermal fluctuations is as
follows. The superconducting fluctuations contribution\cite{Varlamov} to the
voltage has the following form, see dashed lines in Fig.1. It rises very
fast at small currents and gradually decreases to zero when the virtual
Cooper pairs are broken by the electric field. The negative slope at some
point becomes equal or larger that the normal electrons conductivity that is
roughly independent of the transport current (dark green line in Fig.1). The
appearance of the S-shaped I-V curves at certain value of temperature (the
blue curve) is the crossover temperature that will be determined in the
paper.

\begin{figure}[tbp]
\begin{center}
\includegraphics[width=12cm]{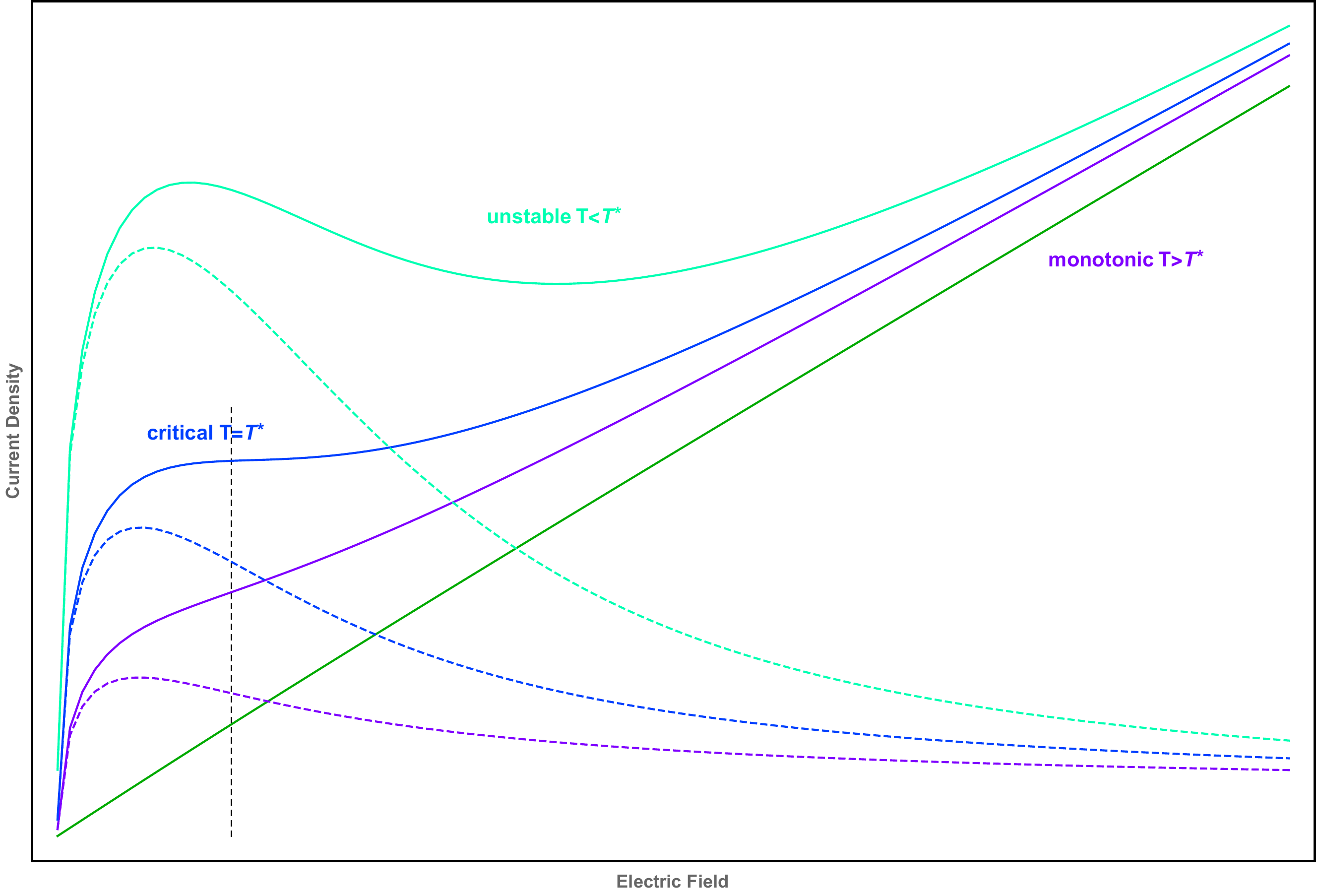}
\end{center}
\par
\vspace{-0.5cm}
\caption{Schematic I-V curves at different temperatures. The dashed lines
are superconducting fluctuation contributions at different temperatures and
the solid lines are total currents contributed both from the normal part
(straight line)and superconducting part. Below transition temperature $%
T^{\ast }$, non-monotonic S shaped I-V curves appear.}
\end{figure}

The rest of the paper is organized as follows. In Section II the time
dependent Ginzburg - Landau model incorporating the effects of thermal
fluctuations is specified, while in Section III the I-V curves are derived
in $D=1,2,3$ within the gaussian approximation. The instability is analyzed
in Section IV by considering a quantum wire experiment and several 2D
materials ranging from thin films to layered superconductors and few atomic
thick new materials. Section V contains conclusions.

\section{Thermal fluctuations and electric field in the time-dependent GL
model}

Unlike in many other second order transitions in condensed matter, some
superconductor - normal transitions exhibit a wide thermal fluctuation
region. Since the discovery of high $T_{c}$ superconductors, the
superconducting fluctuations have been demonstrated to be the prime cause of
many interesting phenomena. For example, fluctuations broaden the critical
region of resistivity in the vicinity of the transition temperature\cite%
{Rullier}, lead to large diamagnetism\cite{Ong} and Nernst effect \cite%
{WangNernst} far above $T_{c}$ etc. The influence is especially enhanced
under strong magnetic fields.

\subsection{The model}

While it is impossible at the present level of our understanding of
superconductivity in these materials to describe the effect of thermal
fluctuations on transport within a microscopic model, the Ginzburg-Landau
(GL) phenomenological description in terms of the order parameter field $%
\Psi $ is a method of choice\cite{Varlamov,Rosenstein} for that purpose. To
describe the thermal fluctuations of the order parameter in $D$ $\,$-
dimensional superconductors a starting point is the GL free energy as a
functional of the order parameter field $\Psi $:%
\begin{equation}
F_{GL}=A\int d^{D}\mathbf{r}\frac{{\hbar }^{2}}{2m^{\ast }}|\mathbf{\nabla }%
\Psi |^{2}+\alpha (T-T_{\Lambda })|\Psi |^{2}+\frac{b}{2}|\Psi |^{4}\text{.}
\label{FGL}
\end{equation}%
For low dimensional superconductors the cross - section ``area" is indeed
area, $A=L_{y}L_{z}$, for $D=1$, while in $D=2$ it is the sample effective
thickness $A=L_{z}$. In the GL potential term, $T_{\Lambda }$ is the
mean-field critical temperature, that can be significantly larger than the
measured critical temperature $T_{c}$ due to strong thermal fluctuations on
the mesoscopic scale\cite{Bu2} and $m^{\ast }$ is effective Cooper pair mass.

For strong fluctuation superconductors away from both the critical range and
the gaussian fluctuations regime at very low temperatures, one have to take
the quartic term in GL free energy into account. The other two parameters $%
\alpha $ and $b$ determine the two characteristic length scales, the
coherence length $\xi ^{2}=\hbar ^{2}/\left( 2m^{\ast }\alpha T_{\Lambda
}\right) $ and the penetration depth $\lambda ^{2}=bc^{2}m^{\ast }/\left(
16\pi e^{2}\alpha T_{\Lambda }\right) $.

The relaxational dynamics and thermal fluctuations of the superconducting
order parameter in the presence of electric field $E$ are conveniently
described by the gauge-invariant time-dependent GL (TDGL) equation\cite%
{Dorsey} with the Langevin white noise:
\begin{equation}
\Gamma _{0}^{-1}\left( \frac{\partial }{\partial \tau }-i\frac{e^{\ast
}\varphi }{\hbar }\right) \Psi =-\frac{1}{A}\frac{\delta F_{GL}}{\delta \Psi
^{\ast }}+\zeta \left( \mathbf{r},\tau \right) \text{.}  \label{TDGL}
\end{equation}%
Here the order parameter relaxation time is given by $\Gamma _{0}^{-1}={%
\hbar }^{2}\gamma /\left( 2m^{\ast }\right) $, where the inverse diffusion
constant $\gamma /2$, controlling the time scale of dynamical processes via
dissipation, is assumed to be real\cite{Dorsey2}. $e^{\ast }=2\left \vert
e\right \vert $. The scalar potential for constant homogeneous electric
field (assume to be applied along the $x$ axis) is $\varphi =-Ex$. The
white-noise forces, which induce the thermodynamical fluctuations, satisfy
the fluctuation-dissipation theorem%
\begin{equation}
\left \langle \zeta ^{\ast }(\mathbf{r},\tau )\zeta (\mathbf{r}^{\prime
},\tau ^{\prime })\right \rangle =\frac{2T}{A\Gamma _{0}}\delta (\mathbf{r}-%
\mathbf{r}^{\prime })\delta (\tau -\tau ^{\prime })\text{.}  \label{noise}
\end{equation}

The electric current density includes two components, $\mathbf{J=J}_{n}+%
\mathbf{J}_{s},$ where $\mathbf{J}_{n}=\sigma _{n}\mathbf{E}$ is the current
density contributed by the Ohmic normal part, and $\mathbf{J}_{s}$ is
fluctuation supercurrent density given by

\begin{equation}
\mathbf{J}_{s}=\frac{ie^{\ast }{\hbar }}{2m^{\ast }}\left( \Psi ^{\ast }%
\mathbf{D}\Psi -\Psi \mathbf{D}\Psi ^{\ast }\right) \text{.}
\label{superphy}
\end{equation}

\subsection{Characteristic scales and dimensionless variables}

In order to facilitate the following discussion and fitting of experimental
I-V curves, let us use characteristic units of \ length, the coherence
length $\xi $, time, the Ginzburg-Landau \textquotedblleft
relaxation\textquotedblright \ time\cite{Rosenstein} $\tau _{GL}=\gamma \xi
^{2}/2$. The order parameter is normalized by $\Psi ^{2}=\left( 2\alpha
T_{\Lambda }/b\right) \psi ^{2}$ and electric field by $E=E_{GL}\mathcal{E}$%
, where
\begin{equation}
E_{GL}=2\hbar /\gamma e^{\ast }\xi ^{3}\text{.}  \label{E_GL}
\end{equation}%
The fluctuation strength is conveniently characterised by the parameter $%
\omega $,
\begin{equation}
\omega =\sqrt{2Gi}\pi \text{,}  \label{omega}
\end{equation}%
related to the $D$ - dimensional Ginzburg number (consistent with the
original definitions in $D=2$) by%
\begin{equation}
Gi_{D}=2\left( \frac{T_{\Lambda }e^{\ast 2}\lambda ^{2}}{Ac^{2}\hbar ^{2}\xi
^{D-2}}\right) ^{2}\text{.}  \label{Gi}
\end{equation}

The TDGL Eq. (\ref{TDGL}), written in dimensionless units reads,
\begin{equation}
\left( D_{\tau }-\frac{1}{2}\mathbf{\nabla }^{2}\right) \psi +\frac{t-1}{2}%
\psi +\left \vert \psi \right \vert ^{2}\psi =\bar{\zeta}\text{,}
\label{dimensionlessTDGL}
\end{equation}%
where $t\equiv T/T_{\Lambda }$, $D_{\tau }=\frac{\partial }{\partial \tau }+i%
\mathcal{E}y$ and $\zeta =\bar{\zeta}\left( 2\alpha T_{\Lambda }\right)
^{3/2}/b^{1/2}$, the white noise correlation takes a dimensionless form:
\begin{equation}
\left \langle \bar{\zeta}^{\ast }(\mathbf{r},\tau )\bar{\zeta}(\mathbf{r}%
^{\prime },\tau ^{\prime })\right \rangle =2\omega t\delta (\mathbf{r}-%
\mathbf{r}^{\prime })\delta (\tau -\tau ^{\prime })\text{.}
\label{noisecorr}
\end{equation}%
Finally, the dimensionless current density $\mathbf{j}_{s}=\mathbf{J}%
_{s}/J_{GL}$, with $J_{GL}=cH_{c2}\xi /2\pi \lambda ^{2}$ as the unit of the
current density, is%
\begin{equation}
\mathbf{j}_{s}=\frac{i}{2}\left( \psi ^{\ast }\mathbf{D}\psi -\psi \mathbf{D}%
\psi ^{\ast }\right) \text{.}  \label{superdimensionless}
\end{equation}%
The problem is clearly nonperturbative, so that one should rely on methods
of a variational nature that are outlined next. The relevant unit of
conductivity is therefore $\sigma _{GL}\equiv J_{GL}/E_{GL}=c^{2}\gamma \xi
^{2}/4\pi \lambda ^{2}$.

\section{The self - consistent approximation calculation of the I-V curve}

A sufficiently simple nonperturbative method is the Hartree - Fock type
self-consistent Gaussian approximation (SCGA)\cite{Dorsey,Bu1,Bu2}. It has
already been applied to other fluctuations phenomena like magnetization\cite%
{Jiang}, Nernst effect\cite{Bu2} and conductivity above $T_{c}$\cite%
{ChenYBCO}.

\subsection{Dynamical gaussian approximation}

The TDGL in the presence of the Langevin white noise, Eq. (\ref%
{dimensionlessTDGL}), is nonlinear, so cannot generally be solved. Since we
will need only the thermal averages of quadratic in $\psi $ quantities, like
the superfluid density and the electric current, a sufficiently simple and
accurate approximation (similar in nature to the Hartree-Fock approximation
in the fermionic models) is the gaussian approximation \cite{Bu1,Bu2,Jiang}.
The nonlinear $\left \vert \psi \right \vert ^{2}\psi $ term in the TDGL Eq.
(\ref{dimensionlessTDGL}) is approximated by a linear one $2\left \langle
\left \vert \psi \right \vert ^{2}\right \rangle \psi $ (there are two
possible contractions between $\psi ^{\ast }$, $\psi $ in $\left \vert \psi
\right \vert ^{2}\psi $, see discussion of this point in \cite{Kovner}):%
\begin{equation}
\left( D_{\tau }-\frac{1}{2}\mathbf{\nabla }^{2}+\frac{t-1}{2}+2\left
\langle \left \vert \psi \right \vert ^{2}\right \rangle \right) \psi (%
\mathbf{r},\tau )=\bar{\zeta}(\mathbf{r},\tau )\text{.}  \label{linearTDGL}
\end{equation}

For stationary homogeneous processes considered here, the superfluid density
$\left \langle \left \vert \psi \right \vert ^{2}\right \rangle $ is just a
constant. Now it takes a form,%
\begin{equation}
\left[ D_{\tau }-\frac{1}{2}\mathbf{\nabla }^{2}+\varepsilon \right] \psi (%
\mathbf{r},\tau )=\bar{\zeta}(\mathbf{r},\tau )\text{,}  \label{Langeven1}
\end{equation}%
where the excitations energy gap\cite{Rosenstein} is,
\begin{equation}
\varepsilon =-\frac{1-t}{2}+2\left \langle \left \vert \psi \right \vert
^{2}\right \rangle \text{.}  \label{Gap}
\end{equation}%
The solution therefore can be written via the Green's function,
\begin{equation}
\psi (\mathbf{r}_{1},\tau _{1})=\int d\mathbf{r}_{2}\int d\tau _{2}G\left(
\mathbf{r}_{1},\tau _{1};\mathbf{r}_{2},\tau _{2}\right) \bar{\zeta}\left(
\mathbf{r}_{2}\mathbf{,}\tau _{2}\right) \text{.}  \label{TDGLsolu}
\end{equation}%
Then the superfluid density, using the noise correlator, Eq.(\ref{noisecorr}%
), can be expressed via the Green's function as,
\begin{equation}
\left \langle \left \vert \psi \left( \mathbf{r}_{1},\tau _{1}\right) \right
\vert ^{2}\right \rangle =2\omega t\int d\mathbf{r}_{2}\int d\tau
_{2}G^{\ast }\left( \mathbf{r}_{1},\tau _{1};\mathbf{r}_{2},\tau _{2}\right)
G\left( \mathbf{r}_{1},\tau _{1};\mathbf{r}_{2},\tau _{2}\right) \text{,}
\label{density}
\end{equation}%
and is a function of the parameter $\varepsilon $ which is determined self
consistently by Eq.(\ref{Gap}).

\subsection{Green's function for a homogeneous constant electric field}

To calculate the response of the system, one needs the well known Green's
function in the presence of electric field:%
\begin{equation}
G\left( \mathbf{r}_{1},\mathbf{r}_{2},\tau \right) =\theta \left( \tau
\right) \frac{1}{\left( 2\pi \tau \right) ^{D/2}}\exp \left[ -\varepsilon
\tau -\mathcal{E}^{2}\frac{\tau ^{3}}{24}-\frac{i\mathcal{E}}{2}\tau \left(
x_{1}+x_{2}\right) -\frac{\left( \mathbf{r}_{1}\mathbf{-r}_{2}\right) ^{2}}{%
2\tau }\right] \text{.}  \label{propagator}
\end{equation}%
The invariance with respect to the time translations is already taken into
account by setting $\tau =\tau _{1}-\tau _{2}$. Using these expressions, the
superfluid density of Eq. (\ref{density}) takes a form,%
\begin{equation}
\left \langle \left \vert \psi \left( \mathbf{r},\tau \right) \right \vert
^{2}\right \rangle =\frac{\omega t}{2^{D-1}\pi ^{D/2}}\int_{0}^{\infty }%
\frac{d\tau }{\tau ^{D/2}}\exp \left[ -2\varepsilon \tau -\mathcal{E}^{2}%
\frac{\tau ^{3}}{12}\right] \text{.}  \label{density1}
\end{equation}%
The integrand in Eq. (\ref{density1}) is divergent as $1/\tau $ when $\tau
\rightarrow 0$ when $D>1$. The cutoff $\tau _{cut}$ is thus required to
account for the inherent UV divergence of the Ginzburg-Landau theory and it
will be addressed below.

Finally the gap equation takes a form%
\begin{equation}
\varepsilon =-\frac{1-t}{2}+\frac{\omega t}{2^{D-2}\pi ^{D/2}}\int_{\tau
_{cut}}^{\infty }\frac{d\tau }{\tau ^{D/2}}\exp \left[ -2\varepsilon \tau -%
\mathcal{E}^{2}\frac{\tau ^{3}}{12}\right] \text{.}  \label{gapeqscaled}
\end{equation}%
After (numerical) solution for the energy gap $\varepsilon $, we turn to
calculation of the supercurrent. While the upper limit of the integration in
Eq.(\ref{gapeqscaled}) is safe (both terms in exponent are positive), the
lower limit (UV) depends on dimensionality.

In Ref. \onlinecite{Bu2}, it was shown that $\tau _{cut}$ in time
dependent Ginzburg Landau and the energy cutoff $\Lambda $ in static
Ginzburg Landau theory are related by%
\begin{equation}
\tau _{cut}=\frac{\hbar ^{2}}{2m^{\ast }\xi ^{2}\Lambda e^{\gamma _{E}}}
\label{cutoff}
\end{equation}%
where $\gamma _{E}$ is Euler constant and $\Lambda $ is the energy cutoff%
\cite{Bu2,Jiang}. Our calculation show that taking value $\tau _{cut}$ from $%
0.1$ to $10$, the physical quantities is essentially unchanged, and is taken
as $\tau _{cut}=1$ in what follows.

\subsection{The electric current density}

The dimensionless supercurrent density along the electric field direction $x$%
, defined by Eq. (\ref{superdimensionless}), expressed via the Green's
functions is
\begin{equation}
\left \langle j_{x}^{s}\right \rangle =i\omega t\int d\mathbf{r}_{2}d\tau
^{\prime }G^{\ast }\left( \mathbf{r}_{1},\mathbf{r}_{2},\tau -\tau ^{\prime
}\right) \frac{\partial }{\partial x}G\left( \mathbf{r}_{1},\mathbf{r}%
_{2},\tau -\tau ^{\prime }\right) +c.c  \label{j1}
\end{equation}%
Performing the integrals, one obtains,%
\begin{equation}
\left \langle j_{x}^{s}\right \rangle =\frac{\omega t\mathcal{E}}{2^{D}\pi
^{D/2}}\int \frac{d\tau }{\tau ^{D/2-1}}\exp \left[ -2\varepsilon \tau -%
\mathcal{E}^{2}\frac{\tau ^{3}}{12}\right] \text{.}  \label{supercurrent}
\end{equation}

Returning to the physical units, the total electric current density reads%
\begin{equation}
J_{x}=E\left \{ \sigma _{n}+\frac{\omega T\sigma _{GL}}{2^{D}\pi
^{D/2}T_{\Lambda }}\int \frac{d\tau }{\tau ^{D/2-1}}\exp \left[
-2\varepsilon \tau -\left( \frac{E}{E_{GL}}\right) ^{2}\frac{\tau ^{3}}{12}%
\right] \right \} \text{,}  \label{j2}
\end{equation}%
where $E_{GL}$ was defined in Eq.(\ref{E_GL}) and the dimensionless
fluctuation stress parameter $\omega $ in Eq. (\ref{omega}). The gap
equation determining the dimensionless energy gap $\varepsilon $ in this
units is

\begin{equation}
\varepsilon =-\frac{1-T/T_{\Lambda }}{2}+\frac{\omega T}{2^{D-2}\pi
^{D/2}T_{\Lambda }}\int \frac{d\tau }{\tau ^{D/2}}\exp \left[ -2\varepsilon
\tau -\left( \frac{E}{E_{GL}}\right) ^{2}\frac{\tau ^{3}}{12}\right] \text{.}
\label{ephys}
\end{equation}

In general there is a factor $k$ relating the two conductivities: $k=\sigma
_{n}/\sigma _{GL}$. The (obtained numerically) value of the energy gap $%
\varepsilon $ should be used. Illustrative results are presented and
compared with experiments in the next section and discussed in the following
one.

\subsection{The dynamical instability point.}

The dynamical instability transition temperature on the phase diagram, $%
T^{\ast }$, see Fig.1, defined as a maximal temperature at which the
instability appears. Mathematically is determined by vanishing of the first
two derivatives, $\frac{dJ_{x}}{dE}=0\ $\ and $\frac{d^{2}J_{x}}{dE^{2}}=0$.
Differentiating the current, Eq.(\ref{j2}) (via chain rule of the gap
equation), results in:%
\begin{eqnarray}
&&\frac{\sigma _{n}T_{\Lambda }}{\sigma _{GL}T^{\ast }}+\frac{\omega }{%
2^{D}\pi ^{D/2}}\int \frac{d\tau }{\tau ^{D/2-1}}\exp \left[ -2\varepsilon
\tau -\left( \frac{E}{E_{GL}}\right) ^{2}\frac{\tau ^{3}}{12}\right]  \notag
\\
&=&E\frac{\omega }{2^{D}\pi ^{D/2}}\int \frac{d\tau }{\tau ^{D/2-2}}\left( 2%
\frac{\partial \varepsilon }{\partial E}+\frac{\tau ^{2}E}{6E_{GL}^{2}}%
\right) \exp \left[ -2\varepsilon \tau -\left( \frac{E}{E_{GL}}\right) ^{2}%
\frac{\tau ^{3}}{12}\right] ;  \label{Eq1}
\end{eqnarray}%
\begin{equation}
\int \frac{d\tau }{\tau ^{D/2-2}}\left \{
\begin{array}{c}
-\frac{E\tau ^{2}}{2E_{GL}}+\frac{E^{3}\tau ^{5}}{36E_{GL}^{3}}+E_{GL}\frac{%
d\varepsilon }{dE}\left( \frac{2E^{2}\tau ^{3}}{3E_{GL}^{2}}-4\right) \\
+4\frac{E\tau }{E_{GL}}\left( E_{GL}\frac{d\varepsilon }{dE}\right)
^{2}-2EE_{GL}\frac{d^{2}\varepsilon }{dE^{2}}%
\end{array}%
\right \} \exp \left[ -2\varepsilon \tau -\left( \frac{E}{E_{GL}}\right) ^{2}%
\frac{\tau ^{3}}{12}\right] =0  \label{Eq2}
\end{equation}%
Together the gap equation (\ref{gapeqscaled}), the dynamical instability
transition temperature $T^{\ast }$ is determined numerically.

\section{Comparison with experiments and discussion}

The results are first applied to a one dimensional superconductors -
metallic wires, and then for several qualitatively different types of 2D
superconductors (as explained above, it is very difficult to observe the
instability phenomenon in purely 3D materials, although in layered high $%
T_{c} $ cuprates close to $T_{c}$ the fluctuations become nearly 3D and the
phenomenon was observed in magnetic field\cite{Tsuei}).

\subsection{I-V curves of 1D Sn nanowires}

We start with 1D nanowires. Granular superconducting $Pb$ and $Sn$ nanowires
of quite regular cross - section and length have been produced by electro -
deposition in nanoporous membranes\cite{1Dexperiment}. It is important to
note that the series of experiments of Ref. \onlinecite{1Dexperiment} on $Pb$
and $Sn$ nanowires is the only one (known to us) in which \textit{both }the%
\textit{\ }current and the voltage drives were employed. This allows a
qualitative understanding of the important difference between the dynamical
behaviour two. We focus on the voltage drive I-V curves of $Sn$.

\begin{figure}[tbp]
\begin{center}
\includegraphics[width=12cm]{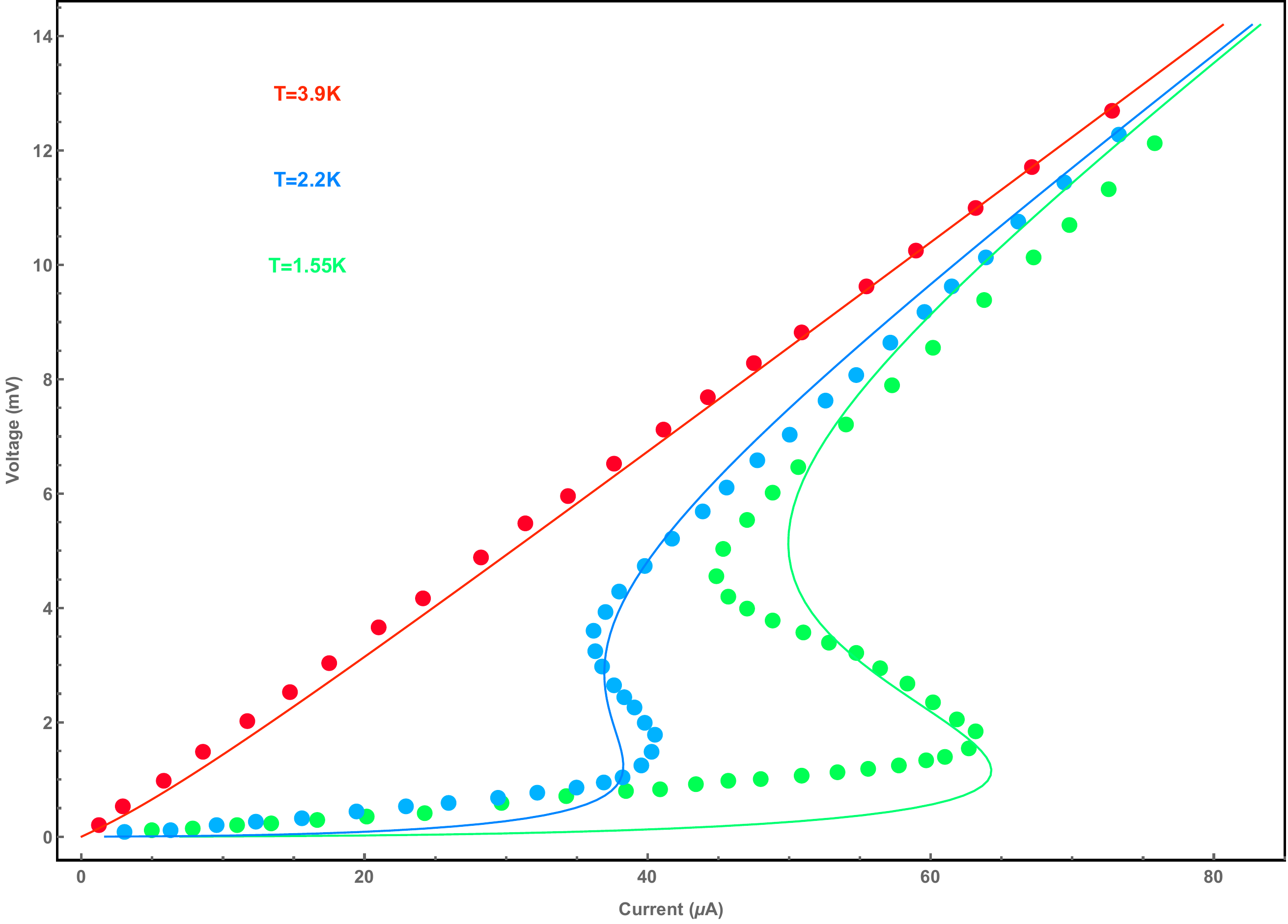}
\end{center}
\par
\vspace{-0.5cm}
\caption{The I-V curves of $1D$ Sn nanowires with different temperature. The
points are the experimental data and the solid lines are the theoretical
results.}
\end{figure}

The I-V curves, measured using the voltage drive at three temperatures, are
shown in Fig. 2 (points). The voltage drive employed clearly demonstrates
the non - monotonic character below the onset $T_{c}\approx T_{\Lambda
}=3.8K $ slightly above the bulk temperature of $Sn$ ($3.72K$). The current
drive experiment on the same sample (see Fig. 3b in Ref. %
\onlinecite{1Dexperiment}) demonstrates the voltage jumps over unstable
domains of the dynamical phase diagrams. The jumps are more pronounced in $Pb
$, see Fig. 3a of Ref. \onlinecite{1Dexperiment}. This is consistent with
the existence of the dynamical instability and was observed in numerous
experiments (see 2D examples below).

The experimental data are fitted by Eqs.(\ref{j2},\ref{ephys}) for $D=1$,
see solid curves. The normal-state conductivity is given, $\sigma
_{n}=3.6\cdot 10^{4}$ $\left( \Omega \ast m\right) ^{-1}$, nanowires are $%
50\mu m$ long with $55nm$ in diameter. Measured material parameters are:
coherence length\cite{Douglass62} $\xi =210nm$, penetration depth $\lambda
=420nm$ and the normal conductivity was obtained from the red doted line in
Fig.2. The value of fitting parameters are: the fluctuation strength
parameter $\omega =0.0043,$ corresponding to the Ginzburg number $%
Gi=9.4\cdot 10^{-7}$, consistent with one dimension Ginzburg number formula,
$Gi_{D=1}=2\left( T_{\Lambda }e^{\ast 2}\lambda ^{2}\xi /Ac^{2}\hbar
^{2}\right) ^{2}\approx 2.9\ast 10^{-7}$ and the conductivity ratio $%
k=\sigma _{n}/\sigma _{GL}=0.08$.

This experiment was already discussed in the framework of TDGL equations
neglecting thermal fluctuations in Ref. \onlinecite{Vodolazov} assuming the
current drive. The dynamical equations were solved numerically and the focus
was on the jumps. It seems to us that the origin of instability cannot
ignore the thermal fluctuations, as explained above. Many more experiments
were performed in 2D.

\subsection{Instability in 2D\protect \bigskip}

Several 2D superconductors exhibit the dynamical instability. We start with
metallic thin films, then proceed to the customary layered materials in
which the coupling between layers is sufficiently small to ensure that
thermal fluctuations dimensionality is $2$. Novel purely 2D materials are
then mentioned. Of course in a 2D superconductor one should measure close to
$T_{c}$ to be able to detect the thermal fluctuations effects like the
dynamical instability. The only possible exceptions are high $T_{c}$
cuprates and novel 2D atomically thick superconductors.

\subsubsection{Thin metallic films near $T_{c}$}

In experiments on $In$ thin films\cite{Ivanchenko} the voltage drive\ was
applied in a narrow temperature range very close to $T_{c}$. For the low
critical temperature superconductor, $T_{\Lambda }$ is very near the bulk
critical temperature. The temperature range (only superconducting states for
$0=1-T/T_{\Lambda }$ $<2.1\%$ are replotted in Fig.3 as dots) nevertheless
is wide enough to exhibit the dynamical instability, for $T/T_{\Lambda
}=0.9821$ and $T/T_{\Lambda }=0.9793$. The coherence length is approximately%
\cite{Brown65} $\xi =300nm$, while the thickness $d$ of $In$ thin films is
ranged\cite{Ivanchenko} from $10nm$ to $300nm$ (less than $\xi $), therefore
in this temperature range coherence length $\xi _{z}>d$ and the thermal
fluctuations are 2D. The normal-state conductivity is $\sigma _{n}=9\ast
10^{4}\left( \Omega \ast cm\right) ^{-1}$. The cross section area
(perpendicular to the current direction) is approximately equal to $%
3.62\times 10^{-7}cm^{2}$ in the fitting, which is consistent with the data
provided in Ref. \onlinecite{Ivanchenko}.

\begin{figure}[tbp]
\begin{center}
\includegraphics[width=12cm]{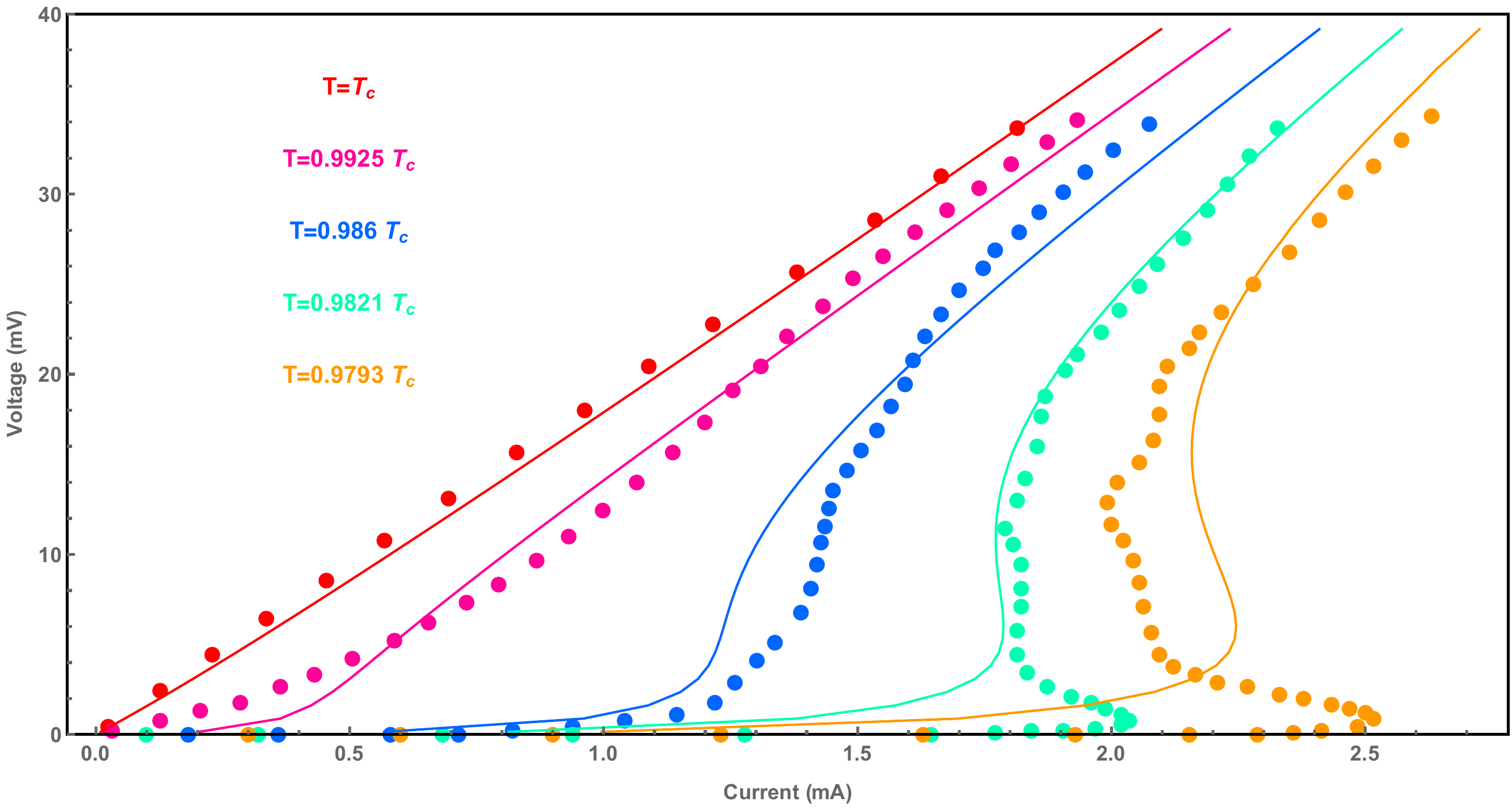}
\end{center}
\par
\vspace{-0.5cm}
\caption{The I-V curves of $In$ thin films\protect \cite{Ivanchenko} and
theoretical fittings at different temperatures.}
\end{figure}

The calculated I-V curves according Eqs.(\ref{j2},\ref{ephys}) for $D=2$,
for different temperature are shown in Fig. 3 as solid curves. The
experimental data are fitted best for the following values of parameters: $%
k=\sigma _{n}/\sigma _{GL}=0.075$ and the fluctuation strength parameter $%
\omega =\sqrt{2Gi}\pi =0.001$ corresponding to the Ginzburg number $%
Gi=5.1\cdot 10^{-8}$, consistent with $Gi_{D=2}=2\left( T_{\Lambda }e^{\ast
2}\lambda ^{2}/L_{z}c^{2}\hbar ^{2}\right) ^{2}\approx 5\cdot 10^{-8}$ with%
\cite{Brown65} $\lambda =296nm$ and\cite{Ivanchenko} $L_{z}=24.1nm$ . The
fit is generally good except very low currents. The reason is obvious:
critical current due to disorder on the mesoscopic scale is not present in
the model.

Sometimes the state close to ``criticality" of the Berezinskii - Kosterlitz -
Thouless variety is theoretically considered as a collection of the bound
vortex - antivortex pairs\cite{Minnhagen}. The critical current clearly seen
in Fig.3 as associated with the pairs ``pinning". In fact in this 2D system
strictly speaking critical current is zero (also seen in data), but it
vanishes exponentially fast as $I\rightarrow 0.$

Much more common superconductors with 2D fluctuations are layered materials
(will be discussed below).

\subsubsection{Layered materials}

Instability in the form of the voltage jumps was observed recently in $%
FeSeTe $ thin film on $Pb(MgNb)TiO$ substrate\cite{Lin15}. Only the current
drive was used, so that the S-shaped I-V curved cannot be determined. Only
the voltage jumps were observed close to $T_{c}$. The thickness of $FeSeTe$
thin films is $200nm$. The layer distance $L_{z}=0.55nm$ \cite{Wu08}. Here,
the normal-state conductivity is taken to be $\sigma _{n}=1.3\cdot
10^{4}\left( \Omega \ast cm\right) ^{-1}$.

\begin{figure}[tbp]
\begin{center}
\includegraphics[width=12cm]{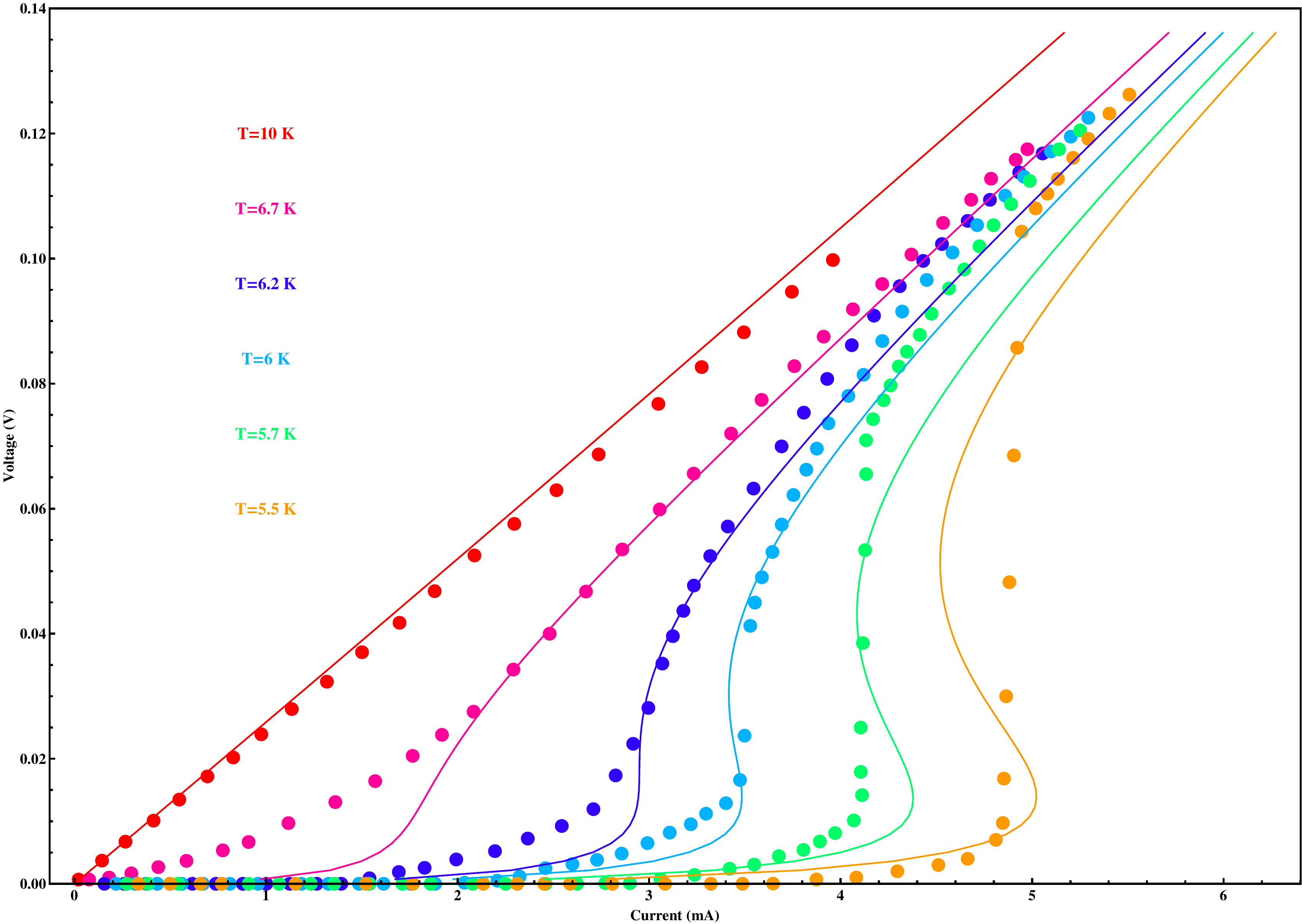}
\end{center}
\par
\vspace{-0.5cm}
\caption{The I-V curves of $FeSeTe$ thin film at different temperatures. The
points are the experimental data and the solid lines are the theoretical
fitting results.}
\end{figure}

The calculated I-V curves of the 2D $FeSeTe$ thin film with different
temperature are shown in Fig. 4 as solid curves. The experimental data of $%
FeSeTe$ in a current driving setup from Ref. \onlinecite{Lin15} are fitted
best for the following values of parameters: $T_{\Lambda }=8K$, $k=\sigma
_{n}/\sigma _{GL}=0.07$ and the fluctuation strength parameter $\omega =%
\sqrt{2Gi}\pi =0.018$ corresponding to the Ginzburg number $Gi=1.6\cdot
10^{-5}$. According to $Gi_{D=2}=2\left( T_{\Lambda }e^{\ast 2}\lambda
^{2}/L_{z}c^{2}\hbar ^{2}\right) ^{2}$, we deduce the sample's effective
penetration depth $\lambda =123.8nm$ (we are not aware of an experimental
determination of the penetration depth from a magnetic measurement).

The I-V curves clearly exhibit a re-entrant behavior for $T<T^{\ast }\approx
6K$. This is hard to observe directly in the current driving experimental
setup. Experiments show that the current driving lead to the ``jump" I-V
Curve and the voltage driving lead to the re-entrant S-Shaped I-V Curve in
the superconducting nanowires at low temperature \cite{Vodolazov}.

\subsubsection{Other layered materials}

The instability in the ultra-thin granular $YBa_{2}Cu_{3}O_{7-\delta }$
nanobridges was clearly observed in a series of works in Ref. \onlinecite{Yeshurun}. Unfortunately a 2D or a 3D model cannot
quantitatively describe these I-V curves since the fluctuations in this
layered material and the temperature range can be described by a more
complicated Lawrence - Doniach model. The generalization is possible but was
not attempted in the present work.

Also, the ``jump" I-V curves in a current driving setup was also reported in
BSCCO\cite{Xiao99} that is clearly 2D. Unfortunately I-V curve at zero
magnetic field (actually field perpendicular to the layers) at one value of
temperature $\left( 76K\text{ for }T_{c}=85.2K\right) $ was measured. As
noted above, the instability has been observed in numerous layered
superconductors under strong magnetic field, but a quantitative
interpretation requires additional parameters describing the magnetic vortex
pinning.

\section{Conclusions}

In this paper, I - V curve of a $D$ dimensional superconductor including the
thermal fluctuations effects is calculated in arbitrary dimension using the
dynamical self consistent gaussian approximation method.\textit{\ }An
unstable region is found when currents flow through superconductor with
temperature below a critical value $T^{\ast }$ at which the I-V curve become
S-shaped. It is shown how the thermal fluctuations generate the instability.
The results are applied to analyse the transport data on various materials
that possess sufficiently strong fluctuations in 1D or 2D. While it is found
that the unstable region can exist also in 3D, the S-Shaped I-V curve in
realistic materials show only in 1D superconductors.

Let us stress that the majority of recent experiments on the resistive state
are performed in the constant current (current driving) regime and at
temperatures close to $T_{c}$. It would be very interesting to observe the
whole S-shaped I-V curve using the \textit{voltage drive} in novel
atomically thick 2D materials as in extensively studied layered ones like $%
BSCCO$.

\textit{Acknowledgments.}

We are grateful to Professor Jian Wang, Professor Guang-Ming Zhang, and Dr.
Ying Xing for valuable discussions. B.R. was supported by NSC of R.O.C.
Grants No. 103-2112-M-009-014-MY3 and  is
grateful to School of Physics of Peking University and Bar Ilan Center for
Superconductivity for hospitality. The work of D.L.  is supported by
National Natural Science Foundation of China (No. 11674007) and is
grateful to NCTS of Taiwan for hospitality.

\end{document}